\input harvmac.tex

\def\nup#1({Nucl.\ Phys.\ $\bf {B#1}$\ (}
\noblackbox
\Title{\vbox{
\hbox{HUTP-97/A105}
\hbox{UCSB-97-25}
\hbox{\tt hep-th/9802016}
}}{Topological Gravity as Large N Topological Gauge Theory}
\bigskip
\centerline{Rajesh Gopakumar $^{(1,2)}$ and  Cumrun Vafa$^{(2)}$}
\bigskip
\centerline{$^{(1)}$ Dept. Of Physics, University of California}
\centerline{Santa Barbara, CA 93106}
\smallskip
\smallskip
\centerline{$^{(2)}$ Lyman Laboratory of Physics}
\centerline{Harvard University}
\centerline{Cambridge, MA 02138}

\vskip .3in

We consider topological closed string theories
on Calabi-Yau manifolds which compute
superpotential terms in the
corresponding compactified type II effective action.
In particular, near
certain singularities we compare the partition function of this
topological theory
(the Kodaira-Spencer theory) to $SU(\infty )$ Chern-Simons theory on the
vanishing
3-cycle.  We find agreement
between these theories, which we check explicitly for the case of
shrinking $S^3$ and Lens spaces, at the perturbative level.
Moreover, the gauge theory
has non-perturbative contributions which have a natural
interpretation in the Type IIB picture.
We provide a heuristic explanation for this
agreement as well as suggest further equivalences
in other topological gravity/gauge systems.

\Date{February 1998}

\lref\strom{A. Strominger, ``Massless Black Holes and Conifolds in String
Theory'', \nup451 ,(1995), 96 .}

\lref\pol{J. Polchinski, ``Dirichlet-Branes and Ramond-Ramond Charges'',
Phys.Rev.Lett. {\bf 75}(1995), 4724.}

\lref\witym{E. Witten, ``Bound States Of Strings And $p$-Branes'',
\nup460 (1996), 335 .}

\lref\strovaf{A. Strominger, C. Vafa, ``Microscopic Origin of the
Bekenstein-Hawking Entropy'', Phys.Lett. {\bf B379} (1996), 99 .}

\lref\narcon{I.Antoniadis, E.Gava, K.S.Narain, T.R.Taylor,
``N=2 Type II- Heterotic duality and Higher derivative
F-terms'', \nup455 (1995), 109.}

\lref\jose{J. Morales, M. Serone, 
``Higher Derivative F-terms in N=2 Strings'',
\nup481 (1996), 389}

\lref\naret{I. Antoniadis, E. Gava, K.S. Narain and T.R. Taylor,
``Topological Amplitudes in String Theory'', \nup413 (1994), 162.}

\lref\BFSS{T. Banks, W. Fischler, S.H. Shenker, L. Susskind,
``M Theory As A Matrix Model: A Conjecture'', Phys.Rev. D55 (1997), 5112 .}

\lref\DVV{R. Dijkgraaf, E. Verlinde, H. Verlinde,
``Matrix String theory'', \nup500 (1997), 43 .}

\lref\BS{T. Banks, N. Seiberg, ``Strings from Matrices'',
\nup497 (1997), 41 .}

\lref\juan{J. Maldacena, ``The Large N Limit of Superconformal Field Theories
and
Supergravity'', hep-th/9711200 .}

\lref\hora{P. Horava, ``M-Theory as a Holographic Field Theory'',
hep-th/9712130 .}

\lref\ghovaf{D. Ghoshal, C. Vafa, ``c=1 String as the Topological Theory of
the Conifold'' , \nup453 (1995), 121 .}

\lref\joao{A. Jevicki, M. Mihailescu, J. P. Nunes,
``Large N WZW Field Theory Of N=2 Strings'',hep-th/9706223 .}

\lref\mart{M. O'Loughlin, ``Chern-Simons from Dirichlet 2-brane instantons'',
Phys.Lett. B385 (1996) 103 .}

\lref\emil{E. Martinec, ``M-theory and N=2 Strings'', 
hep-th/9710122 .}

\lref\witstr{E. Witten -- top.strings.}

\lref\witcs{E. Witten, ``Chern-Simons Gauge Theory As A String Theory'',
hep-th/9207094.}

\lref\vfr{V. Balasubramanian, R. Gopakumar, F. Larsen,
`` Gauge Theory, Geometry and the Large N Limit'', hep-th/9712077 .}

\lref\bcovi{M.Bershadsky, S.Cecotti, H.Ooguri, C.Vafa , ``Holomorphic
Anomalies in Topological Field Theories, \nup405 (1993), 279 .}

\lref\bev{N. Berkovits, C. Vafa,
``N=4 Topological Strings'', \nup433 (1995), 123 .}

\lref\besi{N. Berkovits, W. Siegel ,
``Superspace Effective Actions for 4D Compactifications of
Heterotic and Type II Superstrings'', \nup462 (1996), 213 .}

\lref\ova{ H. Ooguri, C. Vafa, ``Geometry of $N=2$ Strings'',
\nup361 (1991), 469 .}

\lref\ovhe{ H. Ooguri, C. Vafa, ``$N=2$ Heterotic Strings'',
\nup367 (1991), 83 .}

\lref\dv{J. Distler, C. Vafa, ``A Critical Matrix Model at c=1'', Mod. Phys.
Lett. A6, (1991), 259 .}

\lref\wit{E. Witten, ``Quantum Field theory and the Jones Polynomial'',
Commun.Math.Phys.121,(1989) 351 .}

\lref\grossk{D. Gross, I. Klebanov, ``One Dimensional String theory on a
Circle'',
\nup344 (1990), 475 .}

\lref\kazet{V.A.Fateev, V.A.Kazakov, P.B.Wiegmann,
``Principal Chiral Field at Large N'', \nup424 (1994), 505 .}

\lref\mva{S. Mukhi, C. Vafa, ``Two dimensional black-hole as a topological
coset model of c=1 string
   theory'', \nup407 (1993) 667 .}

\lref\gepwit{D. Gepner, E. Witten, ``String Theory on Group Manifolds'',
\nup278 (1986), 493 .}

\lref\lisa{L. Jeffrey, ``Chern-Simons theory on Lens Spaces and Torus Bundles,
and the Semi-classical
Approximation'', Commun.Math.Phys.147, (1992), 563 .}

\lref\roza{L. Rozansky, Commun.Math.Phys. 175 , 275 .}

\lref\wit{E. Witten, ``Quantum Field Theory and the Jones Polynomial'',
Commun.Math.Phys.121, (1989), 351 .}

\lref\peri{V. Periwal, `` Topological Closed-string Interpretation of
Chern-Simons Theory'', Phys. Rev. Lett. {\bf 71} (1993), 1295.}

\lref\ovan{H. Ooguri, C. Vafa, ``All Loop N=2 String Amplitudes'',
\nup451 (1995) 121 .}

\lref\vafcon{C. Vafa, ``A Stringy Test of the Fate of the Conifold''
\nup447 (1995), 252 .}

\lref\wigrr{E. Witten, `` Ground Ring Of Two Dimensional String Theory'',
\nup373 (1992), 187\semi E. Witten and B. Zwiebach, Nucl. Phys.
{\bf B377} (1992) 55.}

\lref\polk{I. Klebanov, A. Polyakov, ``Interaction of Discrete States in
Two-Dimensional String Theory'',
Mod. Phys. Lett. A6 (1991), 3273 .}

\lref\modp{R. Dijkgraaf, G. Moore, R. Plesser, ``The partition function of 2d
string theory'',
\nup394 (1993), 356 .}

\lref\BSV{M. Bershadsky, V. Sadov, C. Vafa, ``D-Strings on D-Manifolds'',
\nup463 (1996), 398 .}

\lref\gopvaf{R. Gopakumar, C. Vafa, ``Branes and Fundamental Groups'',
hep-th/9712048.}

\lref\iz{C. Itzykson, J. Zuber, ``Quantum field theory'', Addison Wesley
Publishing.}

\lref\dilbas{D. Jatkar, B. Peeters, ``String Theory near a Conifold
Singularity'',
Phys.Lett. B362 (1995), 73 .}

\lref\esalb{S. Kachru, E. Silverstein, A. Lawrence, ``On the Matrix Description
of Calabi-Yau Compactifications''
 hep-th/9712223 .}

\lref\bcovii{M. Bershadsky, S. Cecotti,H. Ooguri, C. Vafa ,
``Kodaira-Spencer Theory of Gravity and Exact Results for
Quantum String Amplitudes'', Comm. Math. Phys. {\bf 165} (1994), 311 .}

\newsec{Introduction}

While it has long been suspected that large $N$ gauge theories might
have a string description, it is only recently that physicists have seriously
considered the reverse logic of obtaining gravity (or closed string effects)
from this limit of gauge theories.
That gravity and gauge theory are related in some way is a very
old notion.  The idea of Kaluza-Klein compactifications, in particular,
is a way in which pure gravity leads to gauge theory in lower
dimensions.  This idea and its extensions have been well studied
in the physics literature.
The  reverse idea of obtaining gravity from gauge dynamics
has also been explored, with less clear success.  This includes
attempts at viewing the graviton as a bound state in a gauge theory
(for instance, viewing closed strings as bound states of open strings).

A quite different approach to getting gravity from a gauge theory is
to start with a worldvolume theory as some kind of gauge theory
(or generalisations)
and obtain gravity in the target space
 as a large $N$ limit \BFSS .
One of the main difficulties in checking these proposals is that,
apart from statements protected by supersymmetry, the large $N$ theory
is quite difficult to analyze and it has not
been possible to verify the validity of these conjectures in full detail.

In this paper we initiate a program of studying topological gravity
theories and their connection to large $N$ topological gauge theories
in the context of string theory.
On the one hand, these provide useful and computable testing grounds
for viewing gravity as a large $N$ gauge theory.  On the other hand,
at least some of the topological gravity theories are related
to certain physical
amplitudes in compactified string theories.  The topological theory
computes superpotential
terms in the low energy gravitational action of the resulting
theory.  Thus any proposal
for viewing gravity as a large $N$ gauge theory, such
as the matrix approach \BFSS ,
restricted
to computation of superpotential terms, will have to descend to a
match between a topological gravity theory and a
large $N$ topological gauge theory, of the type we are studying
in this paper.  In this way, we can extract a computable, yet highly
non-trivial
piece out of string theory and see if it has a large $N$
description in the spirit of the matrix approach.

The organization of this paper is as follows:
The topological gravity theories we study arise from topological string
theories and are
reviewed in Sec.2.
In particular, the $N=2$ topological string
computes
perturbative contributions to the $R^2F^{2g-2}$ terms of the type IIB effective
action compactified on Calabi-Yau threefolds, where $R$ is the curvature
and $F$ is the graviphoton field strength (similar
statements hold for the $N=4$ topological strings).
It has been known that near the conifold limit
of Calabi-Yau 3-folds, these amplitudes are related to the
partition function of the  $c=1$ non-critical string at self dual radius. In
Sec.3
we will briefly review this and show how this generalises
to deformations away from the conifold. These turn out to be related to the
amplitudes involving the tachyons and the discrete
states of $c=1$ theory at the self-dual radius.  We also
consider the case of $p$ times the self dual
radius.

Like any string theory the $N=2$ topological string theory has an intrinsically
perturbative definition in terms
of a genus expansion.  Motivated by certain
results in the literature, we propose a large $N$ Chern-Simons theory as a
non-perturbative definition of the ($N=2$) topological string theory (at
least near the point where the Calabi-Yau develops a
vanishing 3-cycle).
In Sec.4 we exactly compute the Chern-Simons partition function on some
3-manifolds which appear as shrinking 3-cycles
in Calabi-Yau compactifications. The $N=\infty$ limit can be taken and we find
that the resulting free energy
reproduces not just the perturbative topological gravity/string theory results
but has additional non-perturbative
pieces. These are seen to be the effects of pair production of wrapped
3-brane-antibrane states near the conifold.
This is something which is not calculable with perturbative string theory and
lends support to the claim of the
$SU(\infty)$ gauge theory being a complete description of the topological
sector of the string theory.
We conclude in Sec.5  by explaining
some aspects of our results along with a discussion of some of the issues
raised in this paper as well as directions for future investigations.

\newsec{Topological String Theories}

In this section we will briefly review the two classes
of topological string theories.  One class corresponds
to the underlying 2d conformal field theory having $N=2$ superconformal
symmetry, as, for example, in a sigma model
on Calabi-Yau threefolds.  The other class corresponds
to theories, which before twisting have a small $N=4$ superconformal
symmetry, such as sigma models on $K3$.  We shall refer
to these two classes as $N=2$ and $N=4$ topological strings respectively.

\subsec{N=2 Topological Strings}
This class of theories was introduced in \ref\witg{E. Witten,
Nucl. Phys. {\bf B340} (1990) 281.}\
and further studied in \witcs ,  \bcovi , \bcovii.
One starts with an $N=2$ superconformal
theory with the two fermionic supercurrents $G^+,G^-$
charged under the $N=2$ $U(1)$ current. After the ``twist'', the spin
assignments
are such that $G^+$ has dimension 1 and $G^-$ has dimension 2.  The structure
of the theory after twisting is identical to that of bosonic strings
where one thinks of $G^+$ as the BRST current $j_{BRST}$ and $G^-$ as the
anti-ghost field
$b$.  Just as is usual in bosonic strings one considers the genus $g$ partition
function
\eqn\fg{F_g=\int_{{\cal M}_g}<\big| G^-\mu_1...G^-\mu_{3g-3}\big|^2>}
where ${\cal M}_g$ denotes the moduli space of genus $g$ surfaces
and $\mu_i$ are the Beltrami differentials (derivatives of the world sheet
metric with respect to moduli).
For the partition function to be non-zero without the insertion
of any further operators, there should be $3g-3$ units of
$U(1)$ charge violation in the measure in order to cancel
the $3-3g$ units of charge coming from the $G^-$'s.  Since an $N=2$
superconformal theory with central charge ${\hat c}$  has a contribution of
${\hat c}(g-1)$ units of charge at genus $g$ (from the measure, due
to twisting) this means that it is for ${\hat c}=3$ that we get the critical
case where the free energy is non-zero for all genera.  This is the case,
in particular, for superconformal theories with
Calabi-Yau threefold target spaces.  Actually, up to conjugation, there
are 2 inequivalent ways to twist an $N=2$ theory, depending
on the relative choice of sign for
left-moving versus right-moving
$U(1)$ currents.  Let us refer to these as $A$ versus $B$ topological
theories.  In the Calabi-Yau realization of these superconformal theories,
the partition functions will depend on the moduli of the Calabi-Yau.
In particular, the A-twisting gives rise to a partition function
which only depends on the Kahler moduli of the Calabi-Yau
(and is subject
to worldsheet instanton corrections) and
arises in questions related to the Coulomb branch of IIA string
compactifications.
The B-twisting depends
only on the complex structure (and is subject to no worldsheet instanton
corrections) and similarly appears in questions
related to the Coulomb branch of IIB
compactifications.
Mirror symmetry exchanges the A/B-twisting of a conformal
theory corresponding to a manifold and its mirror.

The genus zero partition function of topological theories
was studied in detail, beginning with the work \ref\cand{
P. Candelas, X.C. de la Ossa, P.S. Green and L. Parkes,
Nucl. Phys. {\bf B 359} (1991) 21.}
(see \ref\esmi{``Essays in Mirror Symmetry,''
,  S.T. Yau (ed.), International Press 1992. }).
The genus one partition function
was studied in \bcovi\ and was extended to higher genus in \bcovii .
Moreover, it was shown in \bcovii\ that the B-twisted topological
string gives rise to a theory of
topological gravity in 6 dimensions, which was called the
{\it Kodaira-Spencer Theory of Gravity}.
It is a theory of variations of the Calabi-Yau metric through variations of
the complex structure.  The $g$-th loop vacuum
amplitude of the Kodaira-Spencer theory gives rise to the genus
$g$ topological partition function $F_g$.

Let us briefly recall what the Kodaira-Spencer theory is.  Let
$A$ denote an anti-holomorphic 1-form with values in the holomorphic
tangent bundle
of the CY 3-fold.  In other words, the components of it are denoted
by $A_{\bar i}^{j}$.  It is also convenient to define the $(1,2)$-form
$A'$ by
$$A'=A \cdot \Omega\qquad i.e., \qquad A'_{{\bar i}kl}=A_{\bar
i}^j\Omega_{jkl}$$
where $\Omega$ denotes the nowhere vanishing holomorphic threeform on
the CY 3-fold.  Then the Kodaira-Spencer action is defined by
\eqn\ksa{S=\int {1\over 2}A'\partial^{-1}{\overline \partial} A'+
{1\over 6}A'\wedge (A\wedge A)'}
where one restricts $A$ to satisfy the gauge condition $\del A=0$.
Then the equation of motion for $A$
gives
$${\overline \partial} A=[A,A]$$
where the bracket denotes the commutator bracket of $A$'s viewed as vector
fields in the CY 3-fold.  This equation is the Kodaira-Spencer
condition for having a deformation for the ${\overline \partial}$
operator
$${\overline \partial} \rightarrow {\overline \partial}+A\partial$$
satisfying ${\overline \partial}^2=0$.  This equation and its
solution has been extensively studied for Calabi-Yau 3-folds
\ref\tian{G. Tian, in ``Essays on Mirror Manifolds,'',
S.-T. Yau (ed.), International Press 1992. }\ref\anw{A.
Todorov, ``Geometry of Calabi-Yau,'' MPI preprint, 1986.}.

There is a somewhat heuristic argument \bcovii\ which casts the action Eq.\ksa\
into a {\it 3-dimensional} Chern-Simons form. Let us Wick rotate our
complex coordinates $(z_i, \bar{z_i})$ on the Calabi-Yau to a (3,3)
signature. In other words, if $z_i=u_i+iv_i$, then we take $v_i\rightarrow iv_i$
and relabel
$z_i\rightarrow y_i, \bar{z_i}\rightarrow x_i$. The $x_i, y_i$'s are
now like coordinates and momenta. (The Kahler form has become a symplectic
form.)
We can think of the $x_i$'s as
parametrising a (3-dimensional) base $X$ and the $y_i$'s the 3-dimensional
cotangent fibre $Y$. The holomorphic 3-form $\Omega$ is now the volume form on
$Y$ which we can choose to be $\epsilon_{ijk}$ in any patch.

Suppressing the indices denoting the dependence on the base $X$, we can write
the gauge condition (adopting three dimensional vector notation on $Y$)
as $\vec{\nabla}\cdot\vec{A}(\vec{y})=0$ or locally $\vec{A}(\vec{y})
=\vec{\nabla}\times\vec{C}(\vec{y})$. One can then define the symmetric inner
product
\eqn\inprod{Tr[A_1 A_2]=\int_Y\vec{A_1}(\vec{y})\cdot\vec{C_2}(\vec{y})
=\int_Y\vec{A_2}(\vec{y})\cdot\vec{C_1}(\vec{y})}
With this inner product the kinetic term in Eq. \ksa becomes
$$\int_X Tr[A\wedge d_xA]$$
where $A$ here is a 1-form on $X$ with (implicit)
``internal'' vector and coordinate ($y_i$)
indices. We can think of these as group indices for the gauge field
$A$ on $X$. $\vec{A}(\vec{y})$ can be thought of as a generator of
reparametrisation invariance with the gauge condition restricting it
to the group of volume (or $\Omega$) preserving diffeomorphisms of $Y$.

Using the commutator bracket on $Y$,
$$[\vec{A_1},\vec{A_2}]=\vec{\nabla}\times(\vec{A_1}\times\vec{A_2})$$
and the inner product defined above, the cubic term in \ksa\ also goes over
to
$$\int_X Tr[A\wedge A \wedge A].$$
In other words the Kodaira-Spencer action has taken the form of a
Chern-Simons theory with the gauge group of volume preserving
diffeomorphisms on $Y$.

\subsec{Open N=2 Topological Strings}
One can also consider the open string version of the above, as was done
in \witcs .  In the
case of the $A$ twist the boundary condition on the open
string is Dirichlet -- the endpoints are on a collection of
supersymmetric 3-cycles.  Moreover it was shown in \witcs\
that if we take the local model of the Calabi-Yau to be
$T^*M$, where $M$ is a real 3-dimensional manifold, and
take N copies of $M$ as a supersymmetric cycle in $T^*M$, then the
effective theory describing the target space physics is a $U(N)$
Chern-Simons theory on $M$ where the Chern-Simons coupling
$k$ is the string coupling constant $g_s$. The perturbative
expansion of the
Chern-Simons effective action has a perturbative open string
interpretation.
A term in the free energy of the form ${N^h \over k^{2g+h-2}}$ can be
associated with a surface with genus $g$ and $h$ boundary components. This follows from
the 'tHooft identification of Feynman diagrams in the gauge theory with
Riemann surfaces.

The B-version of the twist gives a holomorphic version of
Chern-Simons with Neumann boundary conditions
with strings propagating on the full Calabi-Yau threefold.  This would
correspond to a six dimensional gauge theory.  The corresponding
action is given by \witcs :
\eqn\ach{S=\int \Omega\wedge [{1\over 2} A{\overline \partial}A +{1\over 3}
A\wedge
A\wedge A]}
where $\Omega$ is the holomorphic 3-form of the Calabi-Yau and
$A$ is a $U(N)$ holomorphic gauge connection which is an
anti-holomorphic 1-form with values in the adjoint of $U(N)$ (the
trace over the lie algebra indices is implicit in the above formula).

\subsec{What does the N=2 topological string compute?}
It is natural to ask what the physical meaning of
the topological string amplitudes are, viewed as type II
compactifications on the corresponding Calabi-Yau.  This has
been answered in \bcovii \naret\ where
it was shown that the amplitude $F_g$ computes those
superpotential terms for the $N=2$ theory on the non-compact $R^4$ which have
only string
$g$ loop contributions.
In particular if we consider the type IIB string compactified
on a Calabi-Yau 3-fold, the B version of the twist
computes
corrections to superpotential terms involving vector multiplets
\bcovii \naret. More concretely, these are amplitudes involving $(2g-2)$
graviphotons and 2 gravitons.
The A twisted theory gives similar terms for the hypermultiplet
\naret .  Since the coupling constant of type II strings is
a hypermultiplet (see \besi\ for a more
precise discussion) it follows that the vector multiplet
superpotentials do not receive any non-perturbative quantum
corrections, whereas the hypermultiplet superpotential terms do.

Let us thus concentrate on vector multiplet superpotentials which
are perturbatively exact.  The genus $g$ topological partition function
computes the correction to the effective action in the four dimensional $N=2$
theory of the form
\eqn\core{S=...+ F_g(\{t_i\})\int R^2 F^{2g-2} }
where $F_g$ is the topological genus g amplitude in Eq.\fg\ which
depends on the complex structure
moduli $\{t_i\}$ of the Calabi-Yau (the scalars in the
vector multiplets), $R$ is the Riemann tensor and $F$ is the $N=2$
graviphoton field strength ( the index contraction as well
as the presence of other similar terms is dictated
by supersymmetry).

Thus, each genus computation of the Kodaira-Spencer theory
corresponds to computing different terms in the effective
$N=2$ field theory.  Since in the topological string theory it is natural
to sum these up, it is a natural question to ask how this
can make sense from the field theory perspective.

{}From the correction \core\ it is clear that if we give a
constant expectation value to $F$, i.e. consider
constant $E$ and $B$ fields for the graviphoton
then the partition function of the topological theory
would  compute the correction to the $R^2$ term \narcon\ , namely
$$F(\lambda, t_i)\int R^2$$
where
$$F(\lambda, t_i)=\sum _g\lambda^{2g-2} F_g(\{t_i\}) $$
with $\lambda^2 \sim \langle F^2\rangle $.
To be a little more precise, the $R^2$ term
we have been considering has two relevant contractions
corresponding to the Euler characteristic $\chi$ and the
signature $\sigma$.  The full action is
\eqn\chisig{ S= ...+ {1\over 2}(\chi+{3\over 2}\sigma)F(\lambda, t_i)+{1\over
2}
(\chi-{3\over 2}\sigma){\bar F}({\bar \lambda}, {\bar t_i}) .}
where $\lambda=g_s
(E+iB) $.

We should note
that it is natural in this context to ask if there are
any corrections to the $\int R^2$ term, in the presence of a constant
expectation value to $F^2$, which are not polynomial in $F^2$.
In other words, can we have terms such as ${\rm exp}(-{1\over \lambda})$?
If so this would be the non-perturbative completion of the Kodaira-Spencer
theory.
We will later argue that there are indeed such corrections
and the topological amplitude should be viewed as a function $F(\lambda)$
whose asymptotic expansion for small $\lambda$ gives the above genus $g$
expansion. There would be additional ${\rm exp}(-{1\over \lambda})$ terms,
which are
non-perturbative in the string theory, but are nevertheless topological in
origin.
We will see how the large $N$ Chern-Simons theory
will give us this function $F(\lambda)$
at least near certain singularities.

\subsec{N=4 topological strings}
We can generalize topological $N=2$ strings to topological
$N=4$ strings \bev\ .
If we consider a theory which has $N=4$ superconformal symmetry,
then we can again consider twisting it by choosing an
$N=2$ subalgebra.  In this case there
is a whole sphere's  worth of doing this.   Choose one, and consider
the four supercurrents $G^{\pm},{\tilde G}^{\pm}$.  The N=4 topological
string amplitudes at genus $g$ are defined by
\eqn\Fg{F_g=\int_{{\cal M}_g} \langle
\big| G^-(\mu_1)...G^-(\mu_{3g-3}) \big|^2 \int [ \big|{\tilde
G^+}\big|^2]^{g-1} \int J_LJ_R\rangle}
Note that the net charge violation is $-(2g-2)$ and so an N=4
theory with $\hat c=2$ will give a critical theory.  Examples
include superconformal theories with target spaces $T^4$ or $K3$.
This would be an  Euclidean example.  One could also consider
$N=4$ theories coming from
hyperkahler metrics on $T^*\Sigma$ where $\Sigma$
is a Riemann surface \ova .
If $g\leq 1$ this has a Euclidean signature and if $g\geq 1$ it has a
$(2,2)$ signature.

It turns out that just as the $N=2$ topological string
was modelled after bosonic string theory ($N=0$), the $N=4$
topological string theory is modelled after the $N=2$ string theory.
Note that the critical dimension for the topological
$N=4$ theory is the same as the critical dimension for the
$N=2$ string theory \ova .  This is in  fact
not an accident.  As has been shown in \bev\ the $N=4$ topological
amplitude on a hyperkahler four manifold
 is identical to that of an $N=2$ string propagating on the same manifold!
This in particular means that the corresponding target space gravity
for the $N=4$ topological string is the target space theory of $N=2$ strings
which is known to be self-dual gravity (the analogue of the A and B twists
correspond now to the two distinct ways of
writing the self-dual gravity equations-- the
two ``Heavenly equations'' of Plebanski).

\subsec{Open String Version}
One can also consider the open string version of the $N=4$
topological string, which corresponds
to the open $N=2$ string.  Again there
are two natural boundary conditions.  With Neumann
boundary conditions the effective theory lives
in four dimensions and is self-dual Yang-Mills theory \ref\open{N. Marcus,
Nucl. Phys. {\bf B387} (1992) 263.}\ovhe .
With half Dirichlet and half Neumann conditions, one naturally
considers supersymmetric two cycles and allows the string endpoints
to live on them.  For example, consider the local model for the
four manifold to be $T^*S^2$.  Consider a collection of $N$
$S^2$ cycles in this space.
In this case one obtains by a simple extension
of \ovhe\ the target theory living on $S^2$ and
corresponding to the principal chiral model with group $U(N)$ and
with action
$$S=\int_{S^2} \tr (g^{-1}dg)^2$$
This theory has been extensively studied and is known to be an integrable
model \kazet .

It also
turns out  that, for the $N=4$ topological string the corresponding
gravity theory, i.e. self-dual gravity, can be viewed as the large $N$
limit of the principal chiral model \ref\park{Q-Han Park,
Phys. Lett. {\bf B236} (1990) 429\semi {\bf B238}
(1990) 287.}.  In particular,
self-dual gravity on $T^* S^2$ can be viewed
heuristically as the large $N$ limit of the principal chiral
series on $S^2$. There have also been some checks of this equivalence 
at the quantum level \joao\ .
This structure parallels that of the $N=2$
topological string where the role of volume preserving
diffeomorphisms is now played by $SU(\infty)$ which is the local group of
area preserving diffeomorphisms . 

\subsec{What does the N=4 topological string compute?}

Just as N=2 topological strings compute superpotential
amplitudes in type II string theory compactified
on Calabi-Yau threefolds, something similar is true for $N=4$ amplitudes
\bev .  One can show that they compute
terms in the effective action in six dimensions of the form
$$S= F_g\int R^4 F^{4g-4} +\ldots$$
where $F_g$
is the genus $g$ $N=4$ topological partition function in Eq.\Fg\ .
If we consider the four dimensional space to be $T^{4-k}\times R^{k}$,
$F_g$ is related to computations in toroidal compactifications
on $T^{4-k}$ down to $R^{6+k}$.  The only case where the $F_g$'s have
been computed in detail is for the case where the hyperkahler four
manifold is $R^2\times T^2$ \ovan .
But, just as in the $N=2$ case, one can expect to be able to compute
the $F_g$'s for $K3$ as well (at least, near an $A_n$ singularity).

\newsec{Non-critical Bosonic Strings and $N=2$ Topological Strings}

As reviewed above, the structure of the $N=2$ topological strings
parallels that of bosonic strings.  In fact, that was part
of the motivation for studying them, as they provided
for certain non-critical bosonic string vacua, a topological
string equivalent.
As for the critical $N=2$ topological strings there were a
number of hints \bcovii, \mva, \ref\bwet{B.Gato-Rivera and A.
Semikhatov, Phys. Lett. {\bf B288}(1992) 38.}\ref\bets{
M. Bershadsky, W. Lerche, D. Nemeschansky and N. Warner,
Nucl. Phys. {\bf B401} (1993) 279.}\
that they should get mapped to $c=1$ non-critical strings.
This was in fact established in \ghovaf\ where
it was shown that the $c=1$ non-critical string
corresponding to a conformal field theory on a circle at self-dual radius,
which had been specifically studied in \dv,
is equivalent to the topological $N=2$ theory at the conifold.
This was argued in a number of ways, including the fact that both
were determined by the ground ring, which for $c=1$ at self-dual radius
is given by \wigrr\
$$z_1z_4-z_2z_3=\mu$$
where
\eqn\grgen{z_1=a_La_R\quad z_4=b_Lb_R \quad z_2=a_Lb_R \quad z_3=a_Rb_L}
and $a$ and $b$ denote the basic positive and negative units
of tachyon momenta
and in the $c=1$ terminology $\mu$ has the interpretation of
the cosmological constant.  From the $N=2$ topological string perspective
$z_i$ are the local coordinates describing the Calabi-Yau and $
\mu$ denotes the complex structure of the local model for the Calabi-Yau.
As $\mu \rightarrow 0$, the CY develops a singularity and an $S^3$
shrinks to zero size.  This is known as the conifold singularity.
As further evidence for this identification it was shown
in \ghovaf\ that the genus 0, 1 and 2 results of the Kodaira-Spencer
theory near a conifold, which was studied in \bcovii,
gave results in agreement with the genus 0,1 and 2 partition
function of the $c=1$ string at the self-dual radius.

More evidence for this identification was presented
in \naret :  It was argued in \strom\
that the 3-branes of type IIB theory wrapped around $S^3$
would give rise to a massless hypermultiplet in the limit
of vanishing $S^3$ size.  As a first check of this, it was argued in \strom\
that the genus 0 topological string amplitude is consistent
with such an interpretation.  This check was extended to genus
1 in \vafcon.  More generally it was
shown in a beautiful paper \naret\ that the contribution to
 $R^2F^{2g-2}$ term for a single
hypermultiplet of mass $\mu$
can be computed by a one loop computation (generalizing
the Schwinger computation to the $N=2$ setup)
where the hypermultiplet goes around the loop.
Given the conjectured identification of the conifold
as the locus where we get an extra massless hypermultiplet
\strom\ and the identification of the corresponding
topological string with that of the $c=1$ string at self-dual radius
\ghovaf\ , it was checked that the coefficient $F_g(\mu)$ of $R^2F^{2g-2}$
as a function of $g$ agrees with the genus $g$ partition
function of the $c=1$ string at the self-dual radius.
\foot{An interesting generalisation of this computation to other 
amplitudes in the heterotic side, which reproduce the $c=1$
partition function at arbitrary radius, has been made in 
\jose\ .}   
The partition
function of the $c=1$ string at  self-dual radius is given by
\eqn\eulc{{\cal F}(\mu)=\sum_g \mu^{2-2g} \chi_g}
where $\chi_g$ denotes the Euler characteristic of the moduli
of genus g Riemann surfaces and is given by
$$\chi_g ={B_g\over 2g(2g-2)}$$
where $B_g$ is the g-th Bernoulli number.
This perturbative part is given by the large $\mu$ expansion of
\eqn\fmu{{\cal F}(\mu)=\int^{\infty} {ds\over s^3}e^{-is\mu}
({s/2 \over sinh{s\over 2}})^2}
This expression has
 imaginary terms too, like $e^{-2\pi n\mu}$, which correspond to one of many
possible non-unitary,
non-perturbative completions of the $c=1$ theory.

\subsec{More Detailed Match}
There is more to the $c=1$ non-critical strings than
just the partition function.  In particular, one can consider
the correlation function of physical states.  The states are
described as follows \wigrr, \polk
:  Consider the theory at self-dual radius.  In this case
there is an $SU(2)_L\times SU(2)_R$ current algebra at level one acting
on the left/right-movers.  The physical states can be labeled
by their transformation under this group and it turns out that there is
one copy of each state labeled by
$$|j,m,m'\rangle $$
where $j/2$ denotes the spin under both $SU(2)$'s and $m,m'$ are the $j_3$
quantum numbers, which correspond to left- and right-moving momenta
respectively.  The main question is, what is the interpretation of such
states?  If we add the corresponding operators to the action, it will deform
the theory and thus our question is equivalent to finding the
analog of the $(j,m,m')$ deformations in the type IIB theory near the
conifold.  This is actually relatively straightforward to study.
We started with a theory with a defining equation which can be written
as
$${\rm det}M-\mu=0$$
where
\eqn\Matr{\eqalign{M=\pmatrix{z_1 & z_2 \cr
z_3 & z_4 \cr}}}
Note there is an $SL(2,{\bf C})_L\times SL(2,{\bf C})_R$
action on this manifold given by left/right matrix multiplication
on $M$
$$M\rightarrow K_L \cdot M \cdot K_R$$
Let us consider the most general deformation to the manifold.
It is convenient to write this deformation as
$$z_1z_4-z_2z_3-\mu +\sum_j\epsilon_j(z_1,z_2,z_3,z_4)=0$$
where $\epsilon_j$ is a polynomial in $z_i$ consisting
of monomials of total degree $j$.  Note that the $SL(2)_L\times SL(2)_R$
action being linear in $z_i$ will act on these monomials.  In fact
it is easy to show for infinitesimal deformations (and thus
using $z_1z_4-z_2z_3=\mu$) that they form spin $j/2$ representation of
either $SL(2)$'s.  In particular they are in one to one correspondence
with $|j,m,m'\rangle$, where we identify $m$ and $m'$ with the Cartan
of the respective $SL(2)$'s:
$$U(1)_L:\qquad (z_1,z_2,z_3,z_4)\rightarrow (\alpha z_1,
\alpha z_2, \alpha^{-1} z_3,\alpha^{-1}z_4)$$
$$U(1)_R:\qquad (z_1,z_2,z_3,z_4)\rightarrow (\beta z_1,
\beta^{-1} z_2,\beta z_3,\beta^{-1} z_4)$$
This dictionary allows us to go further in the identification
of Kodaira-Spencer amplitudes and the $c=1$ theory:  We can systematically
add finite deformations away from the conifold limit to give rise to an actual
equation satisfied by a compact Calabi-Yau in a coordinate
patch, and this should be
equivalent
to adding discrete states of the $c=1$ theory at self-dual radius to the
action.  Thus, modulo questions of convergence,
{\it computation of} the $c=1$ {\it partition function for arbitrary finite
deformations by discrete states is equivalent to studying the
partition function for} $N=2$ {\it topological strings for arbitrary
Calabi-Yau} 3-{\it folds}.  The most general amplitude involving $c=1$
discrete states at self-dual radius has not been completely
solved. The class involving tachyon operators has been studied
(see in particular \modp\ and references
therein), and they would correspond to deformations of the defining
equations of the form
$$z_1z_4-z_2z_3+\epsilon(z_1)+\epsilon'(z_4)=\mu$$
where $\epsilon$ and $\epsilon'$ are arbitrary functions of $z_1$ and $z_4$
respectively (we are only writing the deformations to first
order here).

\subsec{$Z_p$ orbifolds of $c=1$}
It is natural to ask what would correspond to changing the radius
in the $c=1$ theory, in the geometrical model.  There is a natural answer to
this,
at least when we consider the radius to be $R=pR_0$ where $p$ is
an integer and $R_0$ is the self-dual radius.  To see this,
note that this can be obtained from the self-dual radius
case by modding out by $Z_p$.  The identification of $z_i$ with
the ground ring generators \grgen\ implies the following action
of $Z_p$ on the ground ring, or equivalently on the 3-fold:
$$(z_1,z_2,z_3,z_4)\rightarrow (\omega z_1 ,z_2,z_3, \omega^{-1}z_4)$$
where $\omega$ is a $p$-th root of unity. On the geometrical
side, modding out the conifold by this symmetry has been considered
(in the same context) in \BSV .  In particular, it was shown
there that it is natural to first rewrite the conifold as
$$z_1z_4=\zeta$$
$$\zeta -\mu=z_2z_3$$
Now modding the first equation by $Z_p$ gives an $A_{p-1}$ type
singularity that can be rewritten in terms
of the invariant variable by defining $u=z_1^p$ and
$v=z_4^p$
$$uv=\zeta^p$$
$$\zeta-\mu=z_2z_3$$
The first equation which described an $A_{p-1}$ singularity
can be deformed to
$$uv=\prod_i(\zeta-\mu_i)$$
$$z_2z_3=\zeta -\mu$$
Physically, this translates into $p$ nearly massless hypermultiplets
with masses $\hat \mu_i=\mu_i-\mu$, corresponding to the $p$
inequivalent $S^3$ cycles, as was discussed extensively in
\BSV .  To map this to the $c=1$ theory at $p$ times the self-dual radius,
we have to decide what the $\hat \mu_i$ are.  Since the
partition function for this theory as a function of the
 `little phase space'
has not been worked out in this case, we will first make a prediction
based on the equivalent geometrical theory:  Since
the physics is dominated by the $p$ light charged
hypermultiplets, corresponding to the D3 brane
wrapped around $p$ inequivalent 3-cycles, the answer will
be the same as what we discussed in the context of the Schwinger
computation of integrating out $p$ light hypermultiplets.  In particular,
the free energy of this theory is expected to be
$$F( \{{\hat \mu_i}\})=\sum_i {\cal F}({\hat  \mu_i} )$$
where ${\cal F}$ denotes the partition function at self-dual radius given in
\fmu .
  Of course we should rewrite the partition function in terms of the
invariant variables, which is simply the $p$ symmetric products
involving ${\hat \mu_i}$.  Could this simple answer be correct?
We will now present evidence in its favor, by showing that
for a simple specific choice of ${\hat \mu_i}$ it reproduces the
$c=1$ partition function at $p$ times the self-dual radius.
The free energy is given by \grossk
$${\partial^2F\over \partial \mu^2}= Re\int_0^{\infty}{ds\over s}
e^{-is\mu}({s/2 \over sinh{s\over 2}})
({sp/2 \over sinh{sp\over 2}}).$$
or equivalently  in an unregularised form \dv
\eqn\cp{\eqalign{F =& \sum_{n,m \in {\bf Z_+}+{1\over 2}}log(n+mp+i\mu) \cr
 =& \sum_{j=1}^{\infty}[\sum_{t=-{p-1\over 2}}^{t={p-1\over 2}}jlog(jp+i\mu
+t)] \cr =& \sum_{t=-{p-1\over 2}}^{t={p-1\over 2}}[\sum_{j=1}^{\infty}
jlog(j+{i\mu +t\over p})] \cr =& \sum_{t=-{p-1\over 2}}^{t={p-1\over 2}}{\cal
F}({\hat \mu_t}),\cr }}
where ${\hat \mu_k}={\mu+ik \over p}$.  Thus the free energy
has the expected structure with $p$ equally spaced
${\hat \mu_k}$.

\subsec{Other Shrinking 3-cycles}

In the geometrical setup we can also consider other 3-cycles shrinking to
zero size. In particular these could be (non-singular) orbifolds of $S^3$.
Examples are given by the Lens spaces $L(p,q)$ and $S^3/G$
where $G$ is an $A-D-E$ discrete subgroup of $SU(2)$.
These do not have any natural conformal field theory interpretation analogous
to the $c=1$ theory
(the modding
out suggested by the geometry will not lead to a modular invariant
conformal theory), and so do not admit a perturbative
description in terms of non-critical bosonic strings.  Nevertheless, we find
that they make sense as far
as the topological string (and IIB compactifications) are concerned and
can compute the corresponding amplitudes.

The lens spaces are quotients of $S^3$,
\eqn\deflens{L(p,q): \qquad  (z_1,z_2) \sim (\omega z_1, \omega^q z_2),\qquad
\omega=e^{2\pi i\over p},
\qquad |z_1|^2 +|z_2|^2 =1. }

Only the  $L(p,0)$ are singular. But that is the case corrresponding to the
$c=1$ string at $p$ times the self dual radius as we have seen from the action
on
the ground ring. The spaces $L(p,1)$ are also special. They correspond to
simple $Z_p$ orbifolds of $S^3$ which happens to be
the same as orbifolding by the cyclic
$A_{p-1}$ discrete subgroup which lies in $SU(2)_L$. One can have $D_n, E_n$
orbifolds of $S^3$ as
well, corresponding to modding out by dihedral and exceptional subgroups of
$SU(2)_L$.

String theory in the vicinity of shrinking cycles such as these, exhibits
some new features compared to the $S^3$ case \gopvaf .
Geometrically, the
important object is the fundamental group of the three cycle. The number of
distinct light hypermultiplets is given by the number of distinct
irreducible representations of the fundamental group. Moreover, the charge
of the hypermultiplet is given by the dimension of the representation.

The Lens spaces $L(p,q)$ have fundamental group $Z_p$ and hence, in string
theory there are $p$ light particles all with $U(1)_{RR}$ charge one and
mass ${\mu \over p}$. The
partition function would be expected to be given by
\eqn\lenz{F_{L(p,q)}=p{\cal F}({\mu \over p})}

In the case of the $D_n, E_n$ orbifolds of $S^3$, the fundamental groups
are the non-abelian dihedral and exceptional groups respectively. This
gives rise to a specific prediction for the charges and number of light
hypermultiplets. We refer the reader to  \gopvaf\ for details.
Accordingly the partition function will reflect this in the various $\mu_i$.
In particular let $F_G$ denote the topological partition function for the case
corresponding to $S^3/G$ where $G$ corresponds to an $A-D-E$ subgroup
of $SU(2)_L$.  Then
\eqn\dyn{F_G(\mu)=\sum_i {\cal F}({a_i \mu \over d})}
where the sum $i$ is over the nodes of the corresponding
affine Dynkin diagram and $a_i$ are the Dynkin indices associated
with the corresponding node, and $d$ is the order of the group, $d=|G|$.

\subsec{Schwinger Pair Creation of Branes}

The physics of the light, wrapped 3-brane in the IIB theory can actually be
described in a
much more familiar manner. It was observed in \narcon\ that the
perturbative contribution to the $R^2F^{2g-2}$ terms is {\it exactly}
the same as
that of a Schwinger type one-loop determinant calculation.
This was done by realizing the computation in a heterotic setup
and doing a one loop computation using heterotic strings.

To spell matters out : The effective action for
the branes in the presence of a constant (self-dual)
electromagnetic graviphoton field would vanish
because of the $N=2$ supersymmetry. However, we can consider a
non-vanishing effective
action with additional  $R^2$ insertions to absorb the fermion zero modes.
Thus the analog of Schwinger's computation of the effective
action for a constant electromagnetic field in the present case
is to study the correction to the $R^2$ term.
To preserve at least half the supersymmetry, the
background field needs to be self dual. In Minkowskian space this means
$$\vec{E}=\pm i\vec{B}.$$

As shown first by Schwinger (see for example \iz\ ),
one can exactly integrate
out the charged field to produce an effective action whose real part
is a polynomial in all (even) powers of the field strength. In the case of
a boson, the one loop determinant reads as
\eqn\schw{\eqalign{F(\vec{E},\vec{B},m) &= {1\over 2}Tr\ln det((i\partial
-eA)^2-m^2) \cr &= {e^2EB\over 2\pi^2}
\int_{\epsilon}^{\infty} {ds\over s^3}e^{-i{sm^2\over 2}}
({s/2 \over sinh{seE\over 2}})
({s/2 \over sin{seB\over 2}}) ,\cr }}
where $E^2-B^2=\vec{E}^2-\vec{B}^2$ and $EB=\vec{E}\cdot\vec{B}$ . In
$QED$, $\epsilon$ is an UV cutoff, which will be replaced in string theory
by the string scale.
Taking a self dual field and redefining $\mu= {m^2\over 2eE}$
we have for the free energy
\eqn\schfn{F(\mu)={e^2E^2\over 2\pi^2}
\int_{\epsilon}^{\infty} {ds\over s^3}e^{-is\mu}
({s/2 \over sinh{s\over 2}})^2 ={e^2E^2\over 2\pi^2}{\cal F}(\mu).}

This has a perturbative expansion (in inverse powers of $\mu$)
which is given in \eulc\ and gives the
higher polynomial corrections to the Maxwell lagrangian. The first couple
of terms in this expansion need an UV cutoff $\epsilon$ --
they diverge as $\mu^2ln({\mu\over \epsilon})$ and $ln({\mu\over\epsilon})$
respectively. In string theory this cutoff will be provided by the string
scale.

The match of \schfn\ with
\eulc\ is thus the standard Schwinger computation
extended to this case. However note that \eulc\
is just an asymptotic expansion of ${\cal F}(\mu)$
valid for large $\mu$.
There is moreover an absorptive part corresponding to pair creation
predicted from \schfn . This
can be evaluated by extending the (imaginary part of the) integral to the
whole real line and closing the contour in the lower half plane to pick up
all the non-zero poles of the $\sinh$ on the negative imaginary axis.
The answer is
\eqn\abs{Im {\cal F}(\mu) = ( )\sum_{n=1}^{\infty}{1\over n^2}e^{-2\pi n\mu}}

In the type IIB context one should also expect such terms: they are naturally
interpreted as the
corrections to the $R^2$ term in
the presence of the constant graviphoton field strength from the pair
production of light wrapped brane-antibrane states. Note, from
Eq. \chisig , that this imaginary part of ${\cal F}(\mu)$ is a correction to
the signature only.
The perturbative (real) part was a correction only to the Euler character.
In
this case, the parameter $\mu={m\over 2E}$. This follows from the formula
$(m=e)$
for the relation between graviphoton charge and mass of the wrapped brane,  
which in turn is proportional to size of the
shrinking 3-cycle (and ${1\over g_s}$) by the BPS condition.

The pair production  is a process that
is not calculable in perturbative string theory but must be computable in
a complete description of the theory. We will see below that an $SU(\infty)$
Chern-Simons
field theory reproduces both these perturbative and non-perturbative
contributions in the case where the shrinking 3-cycles are either $S^3$ or
the Lens spaces. The fact that in the case of $S^3$ the perturbative part of
the Chern-Simons
partition function
at $N=\infty$ agrees with the effective action
near the conifold was already observed in \dilbas\
making use of the results of Periwal on the large N limit
in Chern-Simons theory \peri\ .

In addition to the formal relation between Kodaira-Spencer theory
as volume preserving diffeomorphism Chern-Simons theory,
this was part of the
motivation for our conjecture.  The check we perform for the
lens space is new.

\newsec{Chern-Simons Theory and the $N=\infty$ Limit}

After briefly summarising the results of \wit\
we will outline the
computation of the partition function on $S^3$ and $S^3/Z_p$. Then
we will proceed to take the $N=\infty$ limit and interpret the answer.

\subsec{Chern-Simons Theory}

The Chern-Simons (CS) field theory on an arbitrary 3-dimensional manifold M is
defined as
$$Z[M,N,k] = \int [DA] \exp{[{ik\over 4\pi}\int_M Tr(A\wedge dA+{2\over
3}A\wedge A\wedge A)]}.$$
Here the $N$ refers to the gauge group which we will take to be
$SU(N)$.  The theory is topological in that its definition doesn't
rely on a background metric.
The only other parameter is $k$ -- the level which is
integer quantised -- and plays the role of the (inverse) coupling constant
in the theory.  To compare the partition function with that
of Kodaira-Spencer theory
we will actually need to extend this definition
beyond integral values of $k$ and this can be done in a canonical
way in the cases we are dealing with.  In general this is possible for
manifolds $M$ for which the Chern-Simons partition function does not
involve the ``R''-matrix \ref\wic{E. Witten,
private communication.}.  The manifolds we are dealing
with satisfy this property, and thus have an unambiguous analytic
continuation valid for all $k$.

Moreover, the partition function (or with additional Wilson line insertions)
is often exactly computable \wit. The central
idea is the relation of this three dimensional theory to two dimensional
conformal field theories. This essentially follows from
the Hamiltonian quantisation of
the CS theory on two dimensional ``space'', a Riemann surface $\Sigma$. The
resulting Hilbert space is precisely that of the conformal blocks of the
corresponding current algebra (at level $k$) on the Riemann surface. Since the
Hamiltonian is identically zero, the
partition function on $\Sigma\times S^1$ is, for instance, simply given by the
dimension of this vector space. Thus we can immediately read off
the answer for $T^3$ \peri
$$Z(T^3, N, k)= dim H^{SU(N)}_{T^2}= {(N+k-1)!\over k!(N-1)!}.$$

The process of surgery can be used to relate the partition function of
an arbitrary three manifold to that of a simple one like $S^3$ (possibly
with Wilson lines). The process involves removing a solid torus, about a
curve $C$, from a
manifold $M$ and gluing it back with a diffeomorphism $U\in SL(2,Z)$
on the boundary
$T^2$ to give a topologically new manifold $M^{\prime}$. Then the important
result is
\eqn\surg{Z(M^{\prime})=\sum_j\tilde{U}_{0}^jZ(M; R_j).}
Here, the $R_j$ refers to the insertion of a Wilson line (in representation
$R_j$) along the curve $C$. And $\tilde{U}$ is the representation of $U$
on the genus one Hilbert space -- the space of characters of the level $k$
$SU(N)$ current algebra. This space is labelled by the roots of a finite number
of
representations of $SU(N)$ (shifted by the sum of the positive roots
$\vec{\rho} ={1\over 2}\sum_{+}\vec{\alpha_{+}}$ ). The trivial representation
will be interchangeably labelled as either $\vec{\rho}$ or $0$. A further
necessary datum is that $Z(S^2\times S^1; R_j)=\delta_{j0}$.

Now, $S^3$ in turn can be obtained from gluing two solid
tori along their boundary $T^2$'s (with a twist by the $SL(2,Z)$ matrix $S$
which interchanges the two cycles of the $T^2$). Since gluing two solid
tori with no twist gives $S^2\times S^1$, we have
\eqn\sthree{Z(S^3, N, k)=\sum_j\tilde{S}_{0}^jZ(S^2\times S^1; R_j)
= \tilde{S}_{0,0}}

Thus the main thing we need to know is the action of $SL(2,Z)$ on the
space of characters of $SU(N)$ level $k$ current algebras. This has been
known for a while (See for e.g. \gepwit ) .
They have been cast in a particularly
useful form in \lisa \roza .
The matrix $\tilde{U}$ acting on the Hilbert space $H^{SU(N)}_{T^2}$
associated with the $SL(2,Z)$ matrix
\eqn\umatrix{\eqalign{U= \pmatrix{a & c \cr
b & d \cr}}}
is given by \lisa \roza\
\eqn\tildu{\eqalign{\tilde{U}_{\vec{\alpha},\vec{\beta}}
= {[i sgn(b)]^{N(N-1)/2}\over
(|b|M)^{N-1 \over 2}}e^{[-{i\pi\over
12}(N^2-1)\Phi(U)]}{1\over\sqrt{N}}\times \cr
\sum_{\vec{n} \in {\Lambda^R\over b\Lambda^R}}\sum_{w\in W}(-1)^{|w|}
\exp{{i\pi\over bM}[a\vec{\alpha}^2
-2\vec{\alpha}\cdot(M\vec{n}+w(\vec{\beta}))+
d(M\vec{n}+w(\vec{\beta}))^2]}}}
Here $M\equiv N+k$, $\Lambda^R$ is the root lattice of $SU(N)$ and $W$ the
Weyl Group. $\Phi(U)$ is the Rademacher function
$$\Phi(U)={a+d\over b}-12\sum_{l=1}^{b-1}(({l\over b}))(({ld\over b}))$$
$$((x))=0 \ \ \ x\in {\bf Z}; \qquad = x-[x]-{1\over 2} \ \ otherwise$$

Let us first consider the case of $S^3$. In this case, the partition function
is given by \peri\
\eqn\sphere{\eqalign{Z[S^3,N,k] &= \tilde{S}_{\vec{\rho} ,\vec{\rho}} \cr
&= e^{{i\pi \over 8}N(N-1)}{1\over (N+k)^{N/2}}\sqrt{N+k \over
N}\sum_{w\in W}(-1)^{|w|}\exp{[-{2\pi i\over M}\vec{\rho}\cdot w(\vec{\rho})]}
\cr
&= e^{{i\pi \over 8}N(N-1)}{1\over M^{N/2}}\sqrt{N+k \over
N}\prod_{j=1}^{N-1}\{2\sin({j\pi\over N+k})\}^{N-j}}}
where we have read off from the
above expression for the $SL(2,Z)$ matrix $S$.
We have also carried out, in going
to the third line, a non-trivial sum over the Weyl group.
We note some features of this expression:
$$Z(S^3, N, k=0)=1 $$
$$Z(S^3, N, k)=({k\over N})^{1/2}Z(S^3, k, N)$$
Also, the structure of $Z$ is very reminiscent of that of a matrix model
solved via orthogonal polynomials. These take the general form $(\
)\prod_{j=1}^{N-1}(R_j)^{N-j}$
 which is precisely what we have here with $R_j=2\sin({j\pi\over N+k})$. In
fact, working backwards,
we can write a zero dimensional matrix model which will reproduce this
partition function.

If we examine the asymptotic structure
of the free energy (large but finite $N,k$)
, we can see that it takes the form
$$F(S^3, N, k)= \sum_{g=0, h=1}^{\infty}{k^h\over N^{2g-2+h}}C_{g,h}+
\sum_{g=0}^{\infty}({1\over N^{2g-2}}+{1\over k^{2g-2}})\chi_g $$
where we have assumed that $k<<N$. For $N<<k$, only the first term changes
with the roles of $N$ and $k$ interchanged. This reflects the level-rank
symmetry of the quantum Chern-Simons theory.

\subsec{The Partition Function on $S^3/Z_p$}

We will also be examining the case where the shrinking 3-cycle is
$S^3/Z_p$ . (The reader interested purely in the result of the computation may
skip to Eq. (4.13) .)
This is a special subclass of the spaces $S^3/G$, where $G$ is a
discrete subgroup of $SU(2)_L$ -- in this case, of the A-series. Moreover it
also belongs to a different family of 3-dimensional manifolds, the lens
spaces $L(p,q)$. These are also defined in terms of quotients of $S^3$ as in
\deflens\ .
The space we are considering is $L(p,1)$. It can also be obtained by
surgery from two solid tori with a gluing matrix
\eqn\lens{\eqalign{U = ST^pS = \pmatrix{-1 & 0 \cr
p & -1 \cr}}.}
Thus the partition function is given by
\eqn\lenspart{\eqalign{Z(S^3/Z_p, N, k) &=
(\tilde{S}\tilde{T}^p\tilde{S})_{\vec{\rho},\vec{\rho}} \cr
&=K\sum_{\vec{n} \in {\Lambda^R\over p\Lambda^R}}\sum_{w\in W}(-1)^{|w|}
\exp{-{i\pi\over pM}[M\vec{n}+w(\vec{\rho})+\vec{\rho}]^2}.\cr } }
where $K={1\over \sqrt{N}}{e^{{i\pi\over
4}N(N-1)}\over(Mp)^{N-1\over 2}}e^{[-{i\pi\over
12}(p-1)(N^2-1)]}$. Consider the case where $M\equiv 0 \quad (mod\ p)$. Then
with
$$\vec{n}=\sum_i^{N-1}m_i\vec{\alpha}_i ; \qquad m_i \in \{0,1, \ldots p-1\}$$
we have $\exp{-{i\pi M\over p}<\vec{n},\vec{n}>}=1$ since the $\vec{\alpha}_i$
are the simple
roots which obey $<\vec{\alpha}_i,\vec{\alpha}_j>= 2,1,0$
depending on whether $i=j$,
$i=j\pm1$ or otherwise. So that (using also
$<w(\vec{\rho}),w(\vec{\rho})>=<\vec{\rho},\vec{\rho}>$)
\eqn\simpl{\eqalign{Z(S^3/Z_p, N, k) &=
K\sum_{w\in W}(-1)^{|w|}e^{-{2\pi i\over pM}<w(\vec{\rho}),\vec{\rho}>} \cr
&\times \sum_{\{m_i\}}
\exp{-{2\pi i\over
p}\sum_i^{N-1}m_i[<\vec{\alpha}_i,w(\vec{\rho})>+<\vec{\alpha}_i,\vec{\rho}>]}\cr
&= K\sum_{w\in W}(-1)^{|w|}e^{-{2\pi i\over pM}<w(\vec{\rho}),\vec{\rho}>} \cr
&\times \prod_i^{N-1}(\sum_{m_i=0}^{p-1}\exp{-{2\pi i\over
p}m_i[<\vec{\alpha}_i,w(\vec{\rho})>+<\vec{\alpha}_i,\vec{\rho}>]}). \cr }}
The terms in the product
bracket are each of the form $\sum_m e^{-{2\pi i\over p}mn}$ for
integer $n$. This vanishes unless $n \equiv 0\ (mod\ p)$. In other words, the
product contributes only if
$<w(\vec{\alpha}_i),\vec{\rho}>+<\vec{\alpha}_i,\vec{\rho}>\equiv 0 \
(mod \ p)$ for all $ i = 1\ldots N-1$. ( Here $w\rightarrow w^{-1}$ in the
sum over $w$ and we use the property of the inner product to transfer its
action onto $\vec{\alpha}_i$.) This puts a condition on the elements $w\in W$
that do contribute. To spell it out we need some properties of the roots
and the action of the Weyl Group on them.

The $N-1$ roots lie on a hyperplane in $R^N$. In terms of an orthonormal
basis $\{\vec{e}_i\}$ on $R^N$, the simple roots are given by
$$\vec{\alpha}_i=\vec{e}_i-\vec{e}_{i+1}$$
and the positive roots by ${\bf e}_i-{\bf e}_j$ with $i<j$. Therefore
\eqn\ro{\vec{\rho}={1\over 2}\sum_{k=1}^N(N-2k+1)\vec{e}_k}
The Weyl group
acts simply on the the basis $\{\vec{e}_i\}$ as permutations on the
indices
$$i \rightarrow P_i ; \qquad i, P_i\in 1\ldots N .$$
It is easy then to see that $<\vec{\rho},\vec{\alpha}_i> =1$ and
$<\vec{\rho},w(\vec{\alpha}_i)>=P_{i+1}-P_i$. Thus the permutations that have a
non-zero contribution are those for which
$$P_{i+1}-P_i \equiv -1 \ (mod \ p) .$$

What are these permutations like?  The results depend
sensitively on the value of $k$ mod $p$.  It turns out that
in order to reproduce the gravity answer we need to consider the
the case when $k\equiv 0 \ (mod \ p)$ (the other
limits are discussed in the appendix). Since we are considering
$N=\infty$ limit at the end, and we have considered the
case where $M=N+k=0 \ (mod \ p)$ it implies that we also consider
$N=0 \ (mod\  p)$.
Divide the integers $1\ldots N$
into blocks of size $p$. Consider the permutation on the block $1\ldots p$
\eqn\perm{\eqalign{\pmatrix{1 & 2 & \ldots & p \cr
p & p-1 & \ldots & 1 \cr}}}
together with similar permutations on each of the other blocks. This
clearly satisfies the above condition. Moreover, starting from this
arrangement, we can freely permute the
${N\over p}$ elements, one from each block, which differ from each other by a
multiple of $p$. The resulting permutation also satisfies the condition.
Thus the permutation group that survives of the $S_N$
is $(S_{N\over p})^p$. But we actually have some more allowed
permutations. These are cyclic permutations within  each of the blocks
starting from the one shown in \perm , with the restriction that the cyclic
permutation $C_p$ must be the same in all the blocks.

For each such allowed permutation the term in brackets in the second line
of \simpl\ reduces
to $p$ and we have
\eqn\weylpart{Z(S^3/Z_p, N, k)= Kp^{N-1}
\sum_{w\in W^{\prime}}(-1)^{|w|}e^{-{2\pi i\over
pM}<w(\vec{\rho}),\vec{\rho}>}}
Where $W^{\prime}=C_p\otimes(S_{N\over p})^p$
is the surviving Weyl group. This sum over $W^{\prime}$ actually rather
neatly factorises. If we express $\vec{\rho}$ in \ro\ in terms of
$\vec{\rho}^{\prime}$'s
(where $\vec{\rho}^{\prime}$ is the sum over the positive roots of $SU({N\over
p})$ ), the Weyl sum can be shown to reduce to
\eqn\weylfin{\eqalign{\sum_{c\in C_p}(-1)^{|c|}e^{-({i\pi N\over Mp^2}
\sum_{q=0}^{p-1}(p+2q-1)(p+2C(q)-1)})\times \cr
(\sum_{w\in S_{N\over p}}(-1)^{|w|}e^{-{2\pi p i\over
M}<w(\vec{\rho}^{\prime}),\vec{\rho}^{\prime}>})^p \cr .}}
Thus we get on comparison with Eq. \sphere,
(upto multiplicative factors which are not important for the $N=\infty$
limit)
\eqn\lensfin{Z(S^3/Z_p, N, k)=(\ )Z^p(S^3, N/p, k/p).}

This is an extremely simple, closed form expression which will have a natural
interpretation when
we take the large $N$ limit and compare with the expected result on the
Calabi-Yau side.

%
%

\subsec{The $N=\infty$ limit}

We will be primarily interested in the free energy of the $SU(\infty)$ theory.
Let us take this limit for $S^3$ first. (A similar limit of the exact answer
was studied in \peri\ for the cases of $S^3$
and $T^3$.) Its free energy can be written
\eqn\fsphere{\eqalign{F[S^3,N,k] = lnZ[S^3,N,k] = {N(N-1)\over 2}\ln2 \cr
-{(N-1)\over 2}\ln(k+N) -{1\over 2}\ln N
+\sum_{j=1}^{N-1}jlog[\sin{(k+j)\pi\over N+k}] .\cr}}
Let us rather naively take the  $N=\infty$ limit dropping all the terms
that diverge as positive powers of $N$ (or logarithmically). Then the
contribution that survives comes from the last term which then reads
as
\eqn\finfty{F[S^3,N=\infty,k]=\sum_{j=1}^{\infty}j\log(j+k) .}
At first sight, this might seem too naive. We can adopt a more careful
procedure,
where we take three derivatives with respect to $k$ of Eq.\fsphere\ and then
the
limit. This will ensure that we are dealing with well defined
convergent series. In fact, even the term $\sum_{j=1}^{\infty}j\log(j+k)$ is
divergent as it stands. On doing this it is possible to check that we
recover the result in Eq.\finfty\ or more precisely, its third
derivative
\eqn\delf{{\partial^3F[S^3,N=\infty,k]\over\partial
k^3}=\sum_{j=1}^{\infty}{2j\over(j+k)^3} ={\partial^2\over \partial
k^2}(k\psi(k))}
where $\psi(k)={d\over dk}log \Gamma(k)$ with $\Gamma(k)$ being the usual
Gamma function.
Using the asymptotic Stirling expansion of $\Gamma(k)$, one can also write
(after integrating thrice )
\eqn\fbern{F[S^3,N=\infty,k]={1\over 2}k^2\ln k -{B_1\over 2}ln
k+\sum_{g=2}^{\infty}{(-1)^{g-1}B_g\over 2g(2g-2)}k^{2-2g}.}
This is the same as the genus expansion of the free energy of the $c=1$
Matrix model at the self dual radius provided we make the analytic
continuation $k=i\mu$. As we have seen this is also the
correction to the $R^2$ term in the
perturbative effective action (computed by the topological string)
 near the conifold. This equivalence of the perturbative terms with the $S^3$
free energy was first observed in
\dilbas .

But this was only the large $k$ ($\mu$) asymptotic
expansion of $F[S^3,N=\infty,k]$. The
full partition function contains non-perturbative information as
well. These are terms that go like $e^{-2\pi n\mu}$. To isolate them it helps
to write Eq.\finfty (or the convergent Eq. \delf ) in an integral
representation
\eqn\fint{{\partial^3F[S^3,N=\infty,\mu]\over\partial \mu^3}=\int_0^{\infty}ds
e^{-is\mu}({s/2\over sinh{s\over 2}})^2}

We recognise this as equivalent to Eq. \schfn\ . In other
words the non-perturbative parts are precisely the same as those expected to
compute the contribution of brane-antibrane pair production to the $R^2$
term in the IIB effective action.

We now go onto the case of $S^3/Z_p$ and take the $N=\infty$ limit in \lensfin
, in a
similar manner. Since we had obtained a simple expression in terms of the $S^3$
partition
function, the limit requires no further computation. We see $p$ copies of the
$S^3$
case, but now with mass ${1\over p}$'th what we had before. As discussed
earlier,
this is precisely what we expect when an $S^3/Z_p$ cycle shrinks in a IIB
compactification \gopvaf\ as given in equation \lenz :
there should be $p$ particles becoming light but
with charge ${1\over p}$ of the $S^3$ case. They are distinguished only by
their quantum ${\bf Z_P}$ charges.  As in the $S^3$ case,
the non-perturbative physics of brane-antibrane production
is contained in the exponential terms.
That things work as expected, gives additional support
to the $SU(\infty)$ theory being a full description of the physics
described by $N=2$ topological strings.

\newsec{Heuristic Explanation and Discussion}

We have seen that the $N=\infty$ limit of $SU(N)$ Chern-Simons theory
on $S^3$ or lens spaces reproduces those topological terms in the type IIB
effective action which are
computed by the partition function of the
Kodaira-Spencer theory of gravity in 6 dimensional space corresponding
to the cotangent space $T^* S^3$ or the cotangent of lens space.
As we discussed in section 2 there is a heuristic explanation
\bcovii\ for this match if the Chern-Simons gauge
group were that of {\it volume} preserving diffeomorphisms instead of
$SU(\infty)$
which corresponds to {\it area} preserving diffeomorphisms.  One
explanation for this reduction in the gauge group might be that, since we are
considering
a shrinking 3-cycle, we are at the fixed point of a rescaling
transformation.  If
we rescaled the metric on the whole manifold
by an infinite overall factor it is reasonable to assume that
volume preserving diffeomorphisms contract to area preserving
diffeomorphisms.  This is because we can view $R^3$ roughly
as $S^2$ times a normal direction, and we can view the normal
direction on $S^2$ as being related to the overall rescaling
direction.  Thus, in some sense the $R^3$ shrinks to an $S^2$ and
so volume preserving diffeomorphisms go over to
area preserving diffeomorphisms.  It would be interesting to
make this argument more precise.
One more fact supports this explanation of the reduction
of group:
  If one
considers $T^3$, the topological gravity partition function on $T^*T^3$
is zero.  And the $SU(\infty)$ Chern-Simons theory on $T^3$ though
non-trivial, does not admit a closed string genus expansion \peri .
This correlates with our explanation that the gauge group of
volume preserving diffeomorphisms will
not be reducible to $SU(\infty)$ in such cases, since  $T^3$
is not shrinkable to zero size in a Calabi-Yau threefold.  Note also,
that this suggests that in certain gravity theories we should
not expect a naive $SU(\infty)$ gauge theory to give us
an equivalent system.  In particular, here the relevant
group presumably continues
to be the group of volume preserving diffeomorphisms.
This suggests that, generally,  infinite dimensional gauge groups
(more exotic than
$SU(\infty)$) might be relevant for describing the gravitational
equivalents.

It would also be interesting to compute the large $N$ Chern-Simons
free energy in the case of $S^3/G$ (G = $D,E$ series of subgroups of $SU(2)$)
and compare it with
the expected result \dyn . Perhaps there are also other shrinking 3-cycles that
could be used as test cases.

We have shown that the genus expansion of Kodaira-Spencer theory of gravity
\bcovii\
is only an asymptotic expansion valid for small
string coupling $\lambda$; we saw that there are corrections of the
form ${\rm exp} [-{1\over \lambda}]$.  More generally, we expect
there to be a partition function $F(\lambda)$ valid for {\it all}
$\lambda$.  We saw this to be the  case for $T^* S^3$ and $T^* (S^3/Z_p)$.
We believe this is a general
property. Namely, for every Calabi-Yau threefold we expect to have
a well defined partition function $F(\lambda)$ valid for
all ${\lambda}$, whose asymptotic expansion for small
$\lambda$ reproduces the perturbative topological string expansion
in terms of Riemann surfaces.  It would also be interesting
to understand the origin of  the non-perturbative corrections
in the language of the Kodaira-Spencer theory.  Do they
correspond to some ``topological instantons''?

In this paper we have found a way to relate and compute the partition
function of Kodaira-Spencer theory of gravity in terms of
the Chern-Simons gauge theory for the special case
of a non-compact Calabi-Yau threefold with particular vanishing
3-cycles. In this
context, just as we expect the deformations away from the local
singularity
to be given by amplitudes involving
discrete states in the $c=1$ theory, it will be useful to establish a
similar dictionary with observables in the Chern-Simons theory.

 It would also be very interesting to extend this to the
cases where the Calabi-Yau is compact.   Given that in the
present case we obtained the gravity partition function
by taking a large $N$ Chern-Simons theory on a supersymmetric
3-cycle, the natural guess would be to do a similar
thing in the compact case and consider gauge theories corresponding
to all possible
supersymmetric 3-cycles.  The main puzzle about this
is that naively this ``gauge theory'' is sensitive to $k$
and Kahler classes of Calabi-Yau, whereas we are dealing here with the
complex structure of the Calabi-Yau, on the gravity side.  So,
some kind of mirror symmetry may be at work here.
Another natural gauge theory to consider is the holomorphic version
of Chern-Simons theory in six dimensions \witcs\ whose action
is given in \ach .  It may happen that this theory
at large $N$ computes the B-model topological gravity partition function
on the corresponding Calabi-Yau threefold.
However, since this theory is a gauge theory in six dimensionsit is not easy to
work with,
even though it makes sense since it arises in
topological open string theory.

In the circle of ideas relating the $c=1$ theory, the IIB near the conifold and
the Chern-Simons theory, we
have a reasonable
understanding of the relation between the first two and some understanding of
that between the last two.
But, it will be very pleasing to have some detailed direct understanding of the
connection between
the Chern-Simons theory and the $c=1$ theory. Perhaps, this might be along the
lines of
Douglas' study \ref\doug{M. R. Douglas, ``Chern-Simons-Witten Theory as a
Topological Fermi Liquid '',
 hep-th/9403119}
of large $N$ Chern-Simons theory in terms of free fermions, a representation
familiar in the $c=1$ context.

In this paper we have mainly concentrated on finding a large $N$
gauge description for the Kodaira-Spencer theory of gravity, which corresponds
to $N=2$ topological strings.  As we have briefly indicated,
a lot of the structure is parallel to that for $N=4$ topological
strings.  In this case a natural theory to consider
is the principal chiral model at large $N$ (on $S^2$) and its
relation to self-dual gravity on $T^* S^2$. Another
interesting class to consider is the principal chiral
model on $T^2$.  The large $N$ theory should be equivalent
to $N=4$ topological strings on $T^*T^2=T^2\times R^2$
(note that for this case we do not need any reduction of the group,
so we do not need to have a contractible 2-cycle to apply our considerations).
The interesting point is that
the partition function of this theory has already been computed for all
genera in \ovan.  Given the integrability of principal
chiral models this should lead to an interesting check.
One would also like to consider the large $N$ description
of self-dual gravity in compact situations such as $T^4$
or $K3$.  In this case one conjecture already exists \ovan\
which states that large $N$ holomorphic Yang-Mills theory in
four dimensions may
lead to self-dual gravity.  Another natural guess would be to consider
the large $N$ limit of $U(N)$ self-dual Yang-Mills on $T^4$ and $K3$
and its relation to self-dual gravity on the corresponding spaces.
We also note that the connection between large $N$ $QCD_2$ \ref\grt{D. J.
Gross, W. Taylor, \nup400 (1993),
181},
(in the topological limit)
and $d=2$ topological gravity coupled to topological sigma models
\ref\cmr{S. Cordes, G. Moore, S. Ramgoolam, Commun. Math. Phys. 185 (1997)
543,}
\ref\hor{P. Horava, \nup463 (1996) 238}\ might be viewed as another simple
instance
of the gravity/gauge theory relations we have been considering. Similar 
remarks on the connection between large N gauge theories and (quasi)
topological string theories have been made by Martinec in the context
of N=2 strings \emil\ .  

We have explored some relations between large $N$ topological gauge theories
and
topological gravity/string theory.  This
is very much in the spirit of the matrix conjecture.  In fact,
just as the relevant large $N$ gauge theories in the matrix proposal arise from
the
corresponding open string theory of D-branes,
here the Chern-Simons theory
(or the principal chiral model) are the large $N$ (dirichlet) open string
versions
of the corresponding topological closed string theories.  Moreover, we have
also shown
that the computation we are doing has implications
for physical compactifications of type IIA and type IIB (i.e.
it corresponds to superpotential computations).
The question thus arises as to the precise relation
of our work to the computation of superpotential terms in a matrix theory
approach.
The relation to the usual matrix conjecture is not
very clear.
We have here a bosonic Chern-Simons theory as opposed to a super Yang-Mills.
(It was suggested in \mart\ 
that Chern-Simons theory being
the dirichlet open string version of the A-model, could describe the physics
of instantonic 2-branes in the IIA conifold.)
But then again, we are only computing the superpotential terms.
Thus, perhaps, there is some sense in which the
Chern-Simons describes a sector of a full large N theory. It has been noted
in the recent paper \esalb\
that the matrix description on Calabi-Yau's might be much simpler than on
$T^6$,
in that
gravitational degrees of freedom decouple.
In particular, one is actually dealing with the theory in the vicinity of the
conifold where a
natural T-dual description is in terms of 3-branes. \foot{We also note that an
11-dimensional Chern-Simons like
theory has also been proposed as a candidate M-theory \hora .}

In this context, we must also note that the large $N$ limit that we took is
{\it not}
an 'tHooft like limit.
In the latter, the ratio ${N\over k}$ would have been held fixed. Nevertheless
the
limit we took
was well-defined -- this is similar in spirit to \vfr\ where
some non 'tHooft like large $N$ limits  in Matrix theory were proposed and seen
to give
sensible answers. Again, it would be
interesting to see if the finite $N$ Chern-Simons gauge theory (which was also
computed) has a DLCQ interpretation.
In any case, our toy models might shed some light into many of the hard
questions that face Matrix theory.

\medskip

{\bf Acknowledgements :} We would like to thank V. Balasubramanian,
N. Berkovits, M. Bershadsky, L. Jeffrey,
 A. Lawrence, J. Maldacena, E. Silverstein, A. Strominger
and E. Witten for useful
discussions/correspondence.
The research of R.G. was supported by DOE grant 91-ER40618.
The research of C.V. was supported in part by NSF grant PHY-92-18167.

\bigskip

{\bf Appendix}

In Sec. 4, we evaluated the Lens space partition function for the case
where $N, k \equiv 0\ (mod \ p)$. Here we will state a more general result.
The case when $N+k \equiv 0 \ (mod\  p)$ can be evaluated in quite a similar
manner to that in Sec.4 . The sum over the Weyl group again reduces to that
for $S^3$ but with a  smaller subgroup. The actual group depends on the
congruence properties of $N$. For $N\equiv q \ (mod\  p)$, the final result is
upto a multiplicative factor
\eqn\lensgen{\eqalign{Z(S^3/Z_p, N\equiv q \ (mod\  p), k\equiv -q \ (mod\  p))
\cr = (
)Z^q(S^3,{N-q\over p}+1,{k+q\over p}-1)Z^{p-q}(S^3, {N-q\over p},{k+q\over
p})}}
It is much more difficult to evaluate the partition function in such a
simple closed form for general $N, k$.

It is curious that the large $N$ limit of this expression (which can be
easily taken following the discussion in the text), sensitively depends on
the congruence properties of $k$. (From examining the p=2 case in more
detail, it seems that the limit is independent, as it should be, of how one
takes
$N$ to $\infty$.)

\listrefs

\end

\end